\documentstyle[12pt,aaspp4,epsfig,flushrt]{article}
\singlespace
\newcommand{\cmm}{\,{\rm cm}^{-2}}

\newcommand{\Lya}{Ly$\alpha\ $}

\newcommand{\lunits}{{\rm\,ergs\,s^{-1}\,Hz^{-1}}}

\def\spose#1{\hbox to 0pt{#1\hss}}
\def\lta{\mathrel{\spose{\lower 3pt\hbox{$\mathchar"218$}} \raise 2.0pt\hbox{$\mathchar"13C$}}} 
\def\gta{\mathrel{\spose{\lower 3pt\hbox{$\mathchar"218$}} \raise 2.0pt\hbox{$\mathchar"13E$}}} 

\def\cmm{\,{\rm cm}^{-2}}

\def\lya{Ly$\alpha\ $}

\def\page{\vfill\eject} \def\ni{\noindent}
\def\beq{\begin{equation}}
\def\eeq{\end{equation}}
\def\kmsmpc{\,{\rm km\,s^{-1}\,Mpc^{-1}}}
\def\HI{\hbox{H~$\scriptstyle\rm I\ $}}
\def\HII{\hbox{H~$\scriptstyle\rm II\ $}}

\def\CIV{\hbox{C~$\scriptstyle\rm IV\ $}}
\def\nHI{{\rm HI}}
\def\nH{{\rm H}}

\def\ni{{\noindent}}

\begin{document}
\centerline{\Large {\bf THE INTERGALACTIC MEDIUM}\footnote{A review for the 
Encyclopedia of Astronomy and Astrophysics (Institute of Physics 
Publishing).}}
\bigskip
\author{Piero Madau}
\affil{Institute of Astronomy, Madingley Road, Cambridge CB3 0HA, UK}

\bigskip\bigskip

\ni About half a million years after the Big Bang, the ever-fading cosmic 
blackbody 
radiation cooled below 3000 K and shifted first into the infrared and then 
into the radio, and the smooth baryonic plasma that filled the Universe became 
neutral. The Universe then entered a ``dark age'' which persisted 
until the first cosmic structures collapsed into gravitationally-bound
systems, and evolved into stars, galaxies,
and black holes that lit up the Universe again. Some time between
redshift of 7 and 15, stars within protogalaxies created the first heavy
elements; these systems, together perhaps with an early population of quasars, 
generated the ultraviolet radiation that reheated and reionized the cosmos. 
The history of the Universe during and soon after these crucial formative 
stages is recorded in the all-pervading intergalactic medium (IGM), which is 
believed to contain most of the ordinary baryonic 
material left over from the Big Bang. Throughout the epoch of structure
formation, the IGM becomes clumpy and acquires peculiar motions under the 
influence of gravity, and acts as a source for the gas that gets  
accreted, cools, and forms stars within galaxies, and as a sink for the
metal enriched material, energy, and radiation which they eject.
Observations of absorption lines in quasar spectra at redshifts up to 
5 provide invaluable insight into the chemical composition of the IGM and 
primordial density fluctuation spectrum of some of the earliest formed 
cosmological structures, as well as of the ultraviolet background radiation 
that ionizes them.
 
\page 
\ni {\bf COSMOLOGICAL REIONIZATION}

\ni At epochs corresponding to $z\sim 1000$, the IGM is expected to recombine 
and remain neutral until sources of radiation develop that are capable of 
reionizing it. The detection of transmitted flux shortward of the \lya 
wavelength in the spectra of sources at $z\sim 5$ implies that the hydrogen 
component of this IGM was ionized at even higher redshifts.
There is some evidence that the  double reionization of helium may have 
occurred later, but this is still controversial.
It appears then that substantial sources of ultraviolet photons were already 
present when the Universe was less than 7\% of its current age, perhaps 
quasars and/or young star-forming galaxies: 
an episode of pre-galactic star formation may provide a possible explanation 
for the widespread existence of heavy elements (like carbon, oxygen, and 
silicon) in the IGM, while the integrated radiation emitted from 
quasars is likely responsible for the reionization of intergalactic helium.
Establishing the epoch of reionization and reheating is crucial for determining 
its impact on several key cosmological issues, from
the role reionization plays in allowing protogalactic objects to cool and
make stars, to determining the small-scale structure in the temperature
fluctuations of the cosmic background radiation. Conversely, probing the
reionization epoch may provide a means for constraining competing models for
the formation of cosmic structures, and of detecting the onset of the first
generation of stars, galaxies, and black holes in the Universe.

\bigskip\bigskip {\sf 
\ni INTERGALACTIC HYDROGEN DENSITY 

\ni The proper mean density of hydrogen nuclei at redshift $z$ may be expressed
in standard cosmological terms as:
\begin{eqnarray}
   \bar n_\nH &=& (\rho_{\rm crit}/m_\nH)(1-Y) \Omega_b (1+z)^3  \\
    \nonumber              \\
   &=& (1.1\times 10^{-5}~{\rm cm}^{-3})(1-Y) \Omega_b h^2~(1+z)^3 \; ,
\end{eqnarray}
where $Y$ is the primordial He abundance by mass, $\rho_{\rm crit}=3H_0^2/
(8\pi G)$ is the critical density, $\Omega_b=\rho_b/\rho_{\rm crit}$ is the 
current baryonic density parameter, and $H_0=100\,h\,\kmsmpc$ is the present-day
Hubble constant. Standard nucleosynthesis models together with recent 
observations of 
Deuterium yield $Y=0.247\pm 0.02$, and $\Omega_b h^2=0.0193\pm 0.0014$. 
Thus,
\beq
     \bar n_\nH = (1.6 \times 10^{-7}~{\rm cm}^{-3}) \left({\Omega_b h^2 \over
0.019}\right)(1+z)^3.
\eeq
As some of the baryons had already collapsed into galaxies at $z=2-5$, the 
value of $\Omega_b h^2$ $=0.019$ should strictly be considered as an upper 
limit to the intergalactic density parameter. 
}

\bigskip\bigskip 

\ni Because of the overwhelming abundance of hydrogen, the ionization of 
this element is of great importance for determining the physical state of the 
IGM. Popular cosmological models predict that most of the intergalactic hydrogen
was reionized by the first generation of stars or quasars at $z=7-15$. 
The case that has received the most theoretical studies is one where 
hydrogen is ionized by the absorption of photons, $H+\gamma \rightarrow p+e$ 
(as opposite to collisional ionization $H+e\rightarrow p+e+e$) shortward of 
$912\,$\AA; that is, with energies exceeding $13.6\,$eV, the energy of the 
Lyman edge. The 
process of reionization began as individual sources started to generate 
expanding \HII regions in the surrounding IGM; throughout an \HII region, H is 
ionized and He is either singly or doubly ionized. As more and more sources of 
ultraviolet radiation switched on, the ionized volume grew in size. The 
reionization ended when the cosmological \HII regions overlapped and filled 
the intergalactic space.

\bigskip\bigskip {\sf 
\ni PHOTOIONIZATION EQUILIBRIUM

\ni At every point in a optically thin, pure hydrogen medium of neutral 
density $n_\nHI$, the photoionization rate per unit volume is  
\beq
n_\nHI \int_{\nu_L}^\infty {4\pi J_\nu \sigma_\nH(\nu) \over h_P\nu} d\nu,
\eeq
where $J_\nu$ is the mean intensity of the ionizing radiation  
(in energy units per unit area, time, solid angle, and frequency interval), 
and $h_P$ is the Planck constant. The photoionization cross-section for 
hydrogen in the ground state by photons
with energy $h_P\nu$ (above the threshold $h_P\nu_L=13.6\,$eV)
can be usefully approximated by 
\beq
\sigma_\nH(\nu)=\sigma_L\,(\nu/\nu_L)^{-3},~~~~~~~~~~~~~~~~~\sigma_L=6.3\times
10^{-18}\,{\rm cm^2}.
\eeq 
At equilibrium, this is balanced 
by the rate of radiative recombinations $p+e\rightarrow H+\gamma$ per unit 
volume,
\beq
n_en_p\alpha_A(T),
\eeq
where $n_e$ and $n_p$ are the number densities of electrons and protons, 
and $\alpha_A=\sum\langle \sigma_n v_e\rangle$ is the radiative recombination 
coefficient, i.e. the product of the electron capture cross-section 
$\sigma_n$ and the electron velocity $v_e$, averaged over 
a thermal distribution and summed over all atomic levels $n$. At the commonly 
encountered gas temperature of $10^4\,$K, $\alpha_A=4.2\times 
10^{-13}\,$cm$^3\,$s$^{-1}$.

Consider, as an illustrative example, a point in an intergalactic \HII region 
at (say) $z=6$, with density $\bar n_\nH = (1.6 \times 10^{-7}~{\rm cm}^{-3})
(1+z)^3=5.5\times 10^{-5}\,$cm$^{-3}$. The \HII region surrounds a putative 
quasar with specific luminosity $L_\nu=10^{30}\,(\nu_L/\nu)^2\,\lunits$, and
the point in question is at a distance 
of $r=3\,$Mpc from the quasar. To a first approximation, the mean intensity is 
simply the radiation emitted by the quasar reduced by geometrical dilution,
\beq
4\pi J_\nu={L_\nu\over 4\pi r^2}.
\eeq
We then have for the photoionization timescale:
\beq
t_{\rm ion}=\left[\int_{\nu_L}^\infty {4\pi J_\nu \sigma_\nH(\nu) \over 
h_P\nu} d\nu\right]^{-1}= 5\times 10^{12}\,{\rm s},
\eeq
and for the recombination timescale:
\beq
t_{\rm rec}={1\over n_e\alpha_A}=5\times 10^{16}\,{\rm s}~{\bar n_\nH\over 
n_e}.
\eeq
As in photoionization equilibrium $n_\nHI/t_{\rm ion}=n_p/t_{\rm rec}$, these 
values imply $n_\nHI/n_p\simeq 10^{-4}$, that is, hydrogen is 
very nearly completely ionized.} 

\bigskip\bigskip
\ni A source radiating ultraviolet photons at a finite rate cannot ionize an 
infinite region of space, and 
therefore there must be an outer edge to the ionized volume (this is true 
unless, of course, 
there is a population of UV emitters and all individual \HII regions have 
already overlapped). One fundamental characteristic of the problem is the
very small value of the mean free path for an ionizing photon if the hydrogen
is neutral, $(\sigma_L n_\nH)^{-1}=0.9\,$kpc at threshold, much smaller than 
the radius of the ionized region. If the source spectrum is steep enough
that little energy is carried out by more penetrating, soft X-ray photons, 
we have one nearly completely ionized \HII region, sepated from the outer 
neutral IGM by a thin transition layer or `ionization-front'. 
The inhomogeneity of the IGM is of primary importance for understanding the
ionization history of the Universe, as denser gas recombines faster and is 
therefore ionized at later times than the tenuous gas in underdense regions. 
An approximate way to study the effect of inhomogeneity is to write the 
rate of recombinations as 
\beq
\langle n_en_p\rangle\alpha_A(T)=C\langle n_e\rangle^2 \alpha_A(T)
\eeq
(assuming $T$ is constant in space), where the brackets are the space 
average of the product of the local proton and
electron number densities, and the factor $C>1$ takes into account the degree
of clumpiness of the IGM. If ionized gas with electron density 
$n_e$ density filled uniformly a fraction $1/C$ of the available volume, the 
rest being empty space, the mean square density would be $\langle n_e^2\rangle
=n_e^2/C=\langle n_e\rangle^2C$.  

The IGM is completely reionized when one ionizing photon has been emitted 
for each H atom by the radiation sources, and when the rate of emission of
UV photons per unit (comoving) volume balances the radiative recombination
rate, so that hydrogen atoms are photoionized faster than they can recombine.  
The complete reionization of the Universe manifests itself 
in the absence of an absorption trough in the spectra of galaxies and 
quasars at high redshifts. If the IGM along the line-of-sight to a distant 
source were neutral, the resonant scattering at the wavelength of the \Lya 
($2p\rightarrow 1s; h_P\nu_\alpha =10.2\,$eV) transition of atomic hydrogen
would remove all photons blueward of \lya off the line-of-sight. For any 
reasonable density of the IGM, the scattering optical depth is so large that 
detectable absorption will be produced by relatively small column (or surface) 
densities of intergalactic neutral hydrogen.   

\bigskip\bigskip
{\sf 
\ni GUNN-PETERSON EFFECT 

\ni Consider radiation emitted at some frequency $\nu_e$ that lies blueward  of
\lya by a source at redshift $z_e$, and observed at Earth at frequency 
$\nu_o=\nu_e (1+z_e)^{-1}$. At a redshift $z$ such that 
$(1+z)=(1+z_e)\nu_\alpha/\nu_e$, the emitted photons pass through the local 
\Lya resonance as they propagates towards us through a smoothly distributed
sea of neutral hydrogen atoms,  
and are scattered off the line-of-sight with a cross-section
(neglecting stimulated emission) of 
\beq
\sigma[\nu_o(1+z)]={\pi e^2 \over m_e c}f\, \phi[\nu_o(1+z)],
\label{eq:cross}
\eeq
where $f=0.4162$ is the upward oscillator strength for the
transition, $\phi$ is the line profile function [with normalization $\int 
\phi(\nu)d\nu=1$], $c$ is the speed of light, 
and $e$ and $m_e$ are the electron charge and mass, respectively.
The total optical depth for resonant scattering at the observed frequency is
given by the line integral of this cross-section times the neutral 
hydrogen proper density $n_\nHI(z)$,
\beq
\tau_{\rm GP}=\int_0^{z_e} \sigma[\nu_o(1+z)] n_\nHI(z) {d\ell\over dz}dz,
\eeq
where $d\ell/dz=c H_0^{-1} (1+z)^{-1} [\Omega_M(1+z)^3+\Omega_K(1+z)^2+
\Omega_\Lambda]^{-1/2}$ is the proper line element in a 
Friedmann-Robertson-Walker metric, and
$\Omega_M$, $\Omega_\Lambda$, and $\Omega_K=1-\Omega_M-\Omega_{\Lambda}$
are the matter, vacuum, and curvature contribution to
the present density parameter. 
As the scattering cross-section is sharply peaked around 
$\nu_\alpha$, we can write
\beq
\tau_{\rm GP}(z)=\left({\pi e^2 f \over m_e c \nu_\alpha}\right)\, 
{n_\nHI\over (1+z)}\, {d\ell\over dz}.
\eeq
In an Einstein-de Sitter ($\Omega_M=1$, $\Omega_\Lambda=0$) Universe, this
becomes
\beq
\tau_{\rm GP}(z)={\pi e^2 f \over m_e H_0 \nu_\alpha}\, {n_\nHI\over 
(1+z)^{3/2}}=
6.6\times 10^3 h^{-1} \left(\frac {\Omega_b h^2} {0.019} \right) 
\,{n_\nHI\over \bar n_\nH} \,(1+z)^{3/2}. \label{eq:GP}
\eeq
The same
expression for the opacity is also valid in the case of optically thin (to
\lya scattering) discrete clouds as long as $n_\nHI$ is replaced with
the average neutral density of individual clouds times their volume filling
factor.
}

\bigskip \bigskip

\ni In an expanding Universe homogeneously filled with neutral hydrogen,
the above equations
apply to all parts of the source spectrum to the blue of \Lya. An absorption 
trough should then be detected in the level of the rest-frame
UV continuum of the quasar; this is the so-called ``Gunn-Peterson effect''. 
Between the discrete absorption lines of the \Lya
forest clouds, quasar spectra do not show a pronounced Gunn-Peterson 
absorption trough. The current upper limit at $z_e\approx 5$ is $\tau_{\rm GP}
<0.1$ in
the region of minimum opacity, implying from equation (\ref{eq:GP}) a neutral
fraction of $n_\nHI/\bar n_\nH<10^{-6}\,h$. Even if 99\% of all the cosmic 
baryons fragment at these epochs into structures that can be identified with 
quasar absorption systems, with only 1\% remaining in a smoothly distributed
component, the implication is a diffuse IGM which is
ionized to better than 1 part in $10^4$. 

In modern interpretations of the IGM, it is difficult to use the Gunn-Peterson 
effect to quantify the amount of ionizing radiation that is necessary  to keep
the neutral hydrogen absorption below the detection limits. This is because,
in hierarchical clustering scenarios for the formation of cosmic structures
(the Cold Dark Matter model being the most studied example), the 
accumulation of matter in overdense regions under the influence of gravity
reduces the optical depth for \lya scattering considerably below the average
in most of the volume of the Universe, and regions of minimum opacity occur
in the most underdense areas (expanding `cosmic minivoids').  

\bigskip\bigskip

\ni {\bf A CLUMPY IGM}

\ni Owing to the non-linear collapse of cosmic structures, the IGM is 
well known to be highly inhomogeneous. The discrete gaseous systems detected 
in absorption in the spectra of high-redshift quasars blueward of the \lya 
emission line are assigned different names based on the appearance of their 
absorption features (see Figure 1). 

The term ``\lya forest'' is used to denote the plethora of narrow absorption
lines whose measured equivalent widths imply \HI column densities ranging from 
$10^{16}\,$cm$^{-2}$ down to $10^{12}\,$cm$^{-2}$. These systems,  observed to 
evolve rapidly with redshift between $2<z<4$, have 
traditionally been interpreted as intergalactic gas clouds associated with 
the era of baryonic 
infall and galaxy formation, photoionized (to less than a
neutral atom in $10^4$) and photoheated (to temperatures close to 20,000 K)
by a ultraviolet background close to the one inferred from the integrated 
contribution from quasars. Recent spectra at high resolution and 
high signal-to-noise obtained with the {\it Keck} telescope have shown that 
most \lya forest clouds at $z\sim 3$ down to the detection limit of the 
data have undergone some chemical enrichment, as evidenced by weak, but 
measurable \CIV lines. The typical inferred metallicities range from 
0.3\% to 1\% of solar values, subject to uncertainties of photoionization
models. Clearly, these metals were produced in stars that formed in a denser
environmemt; the metal-enriched gas was then expelled from the regions of star
formation into the IGM. 

An intervening absorber at redshift $z$ having a neutral hydrogen column 
density exceeding $2\times 10^{17}\,$cm$^{-2}$ is optically thick to 
photons having energy greater than 13.6 eV, and produces a discontinuity at 
the hydrogen Lyman limit, 
i.e. at an observed wavelength of $912\,(1+z)$\AA. These scarcer ``Lyman-limit 
systems'' (LLS) are associated with the extended gaseous haloes of bright 
galaxies near the line-of-sight, and have metallicities which appear to be 
similar to that of \lya forest clouds.

In ``damped \lya systems' the \HI
column is so large ($N_\nHI\gta 10^{20}\,$cm$^{-2}$, comparable to the 
interstellar surface density of spiral galaxies today) that the radiation 
damping wings of the \lya line profile become detectable. While relatively 
rare, damped systems account for  
most of the neutral hydrogen seen at high redshifts. The typical metallicities 
are about 10\% of solar, and do not evolve significantly over a redshift
interval $0.5<z<4$ during which most of today's stars were actually formed.

Except at the highest column densities,
discrete absorbers are inferred to be strongly photoionized. From quasar
absorption studies we also know that neutral hydrogen accounts for only a
small fraction, $\sim 10\%$, of the nucleosynthetic baryons at early epochs.

\begin{figure}
\psfig{file=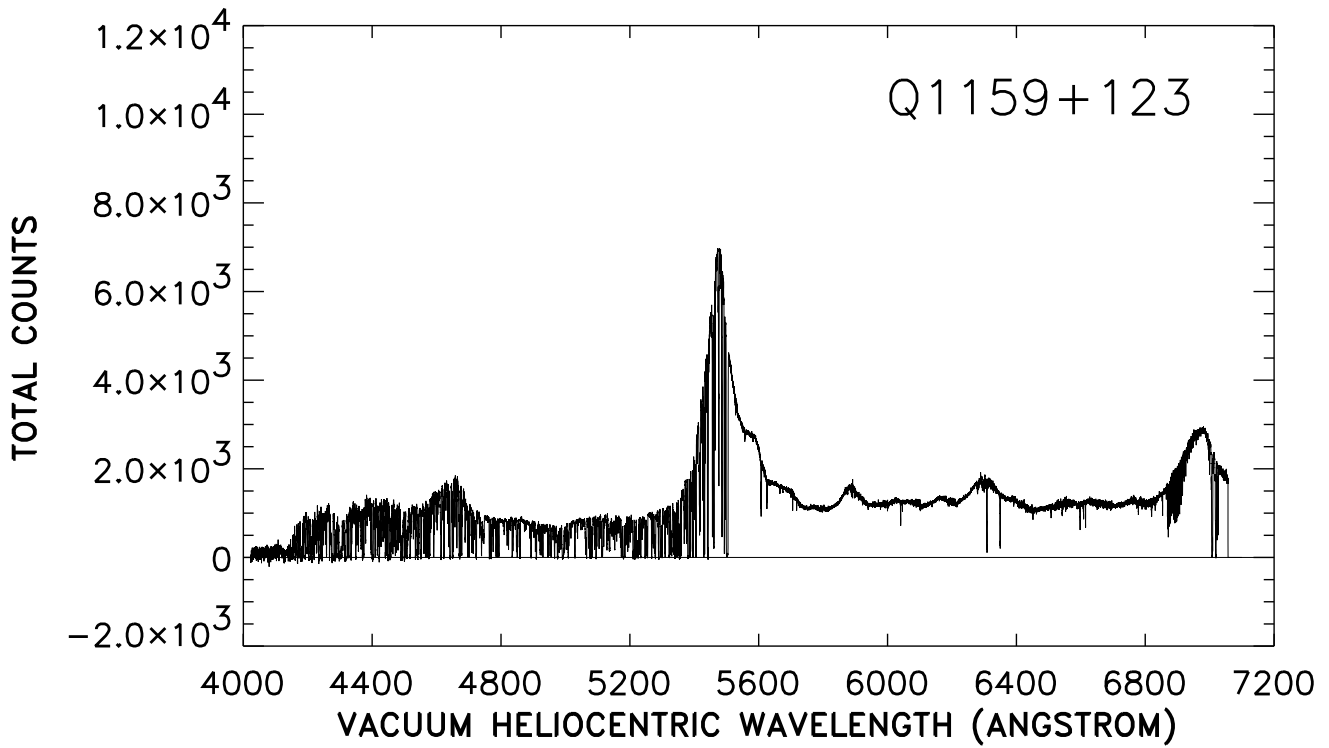,angle=0,width=1\textwidth}
\caption{\footnotesize High resolution ($\lambda/\Delta \lambda=37,0000$) 
spectrum of the 
$z_e=3.50$ quasar Q1159+123, taken with the {\it Keck} High Resolution 
Spectrograph (exposure time 8 h). The data are taken from Songaila (1998). 
The \lya forest is clearly seen in absorption blueward of the atomic hydrogen 
\lya emission line from the quasar (the broad peak at 
$5470\,$\AA), and is produced by resonant \lya scattering in gas clouds along 
the line-of-sight between us and the quasar. 
The spectrum shows a Lyman-limit system just shortward of $4150\,$\AA, 
i.e. 
close to the quasar emission redshift. Most of the features between the \lya
and \CIV emission (the other broad peak just below $7000\,$\AA) are \CIV 
intergalactic absorption lines.  
\label{fig1}}
\end{figure}

\bigskip\bigskip
{\sf 
\ni DISTRIBUTION OF COLUMN DENSITIES AND EVOLUTION

\ni The bivariate distribution $f(N_\nHI,z)$ of \HI column densities and redshifts
is defined by the probability $dP$ that a line-of-sight intersects a cloud 
with column density $N_\nHI$ in the range $dN_\nHI$, at redshift $z$ in the 
range $dz$,
\beq
dP=f(N_\nHI,z)dN_\nHI dz.
\eeq
As a function of column, a single power-law with slope $-1.5$  appears to 
provide at high redshift a surprisingly good description over 9 
decades in $N_\nHI$, i.e. from $10^{12}$ to $10^{21}\,\cmm$. 
It is a reasonable approximation to use for the distribution 
of absorbers along the line-of-sight:
\begin{equation}
f(N_\nHI,z)=A\,N_\nHI^{-1.5}(1+z)^{\gamma}. \label{eq:dis}
\end{equation}
\lya forest clouds and Lyman-limit systems appear to evolve at slightly 
different rates, with $\gamma=1.5\pm 0.4$ for the LLS and
$\gamma=2.8\pm 0.7$ for the forest lines. Let us assume, for 
simplicity, a single redshift exponent, $\gamma=2$, for the entire range in
column densities. In the power-law model (\ref{eq:dis}) the number $N$ of 
absorbers with columns greater than $N_\nHI$ per unit increment of redshift is
\beq
{dN\over dz}=\int_{N_\nHI}^\infty f(N_\nHI',z)dN_\nHI'=2A N_\nHI^{-0.5}(1+z)^2.
\eeq
A normalization value of $A=4.0\times 10^7$ produces then $\sim 3$ LLS per 
unit redshift at $z=3$, and, at the same epoch, $\sim 150$ forest lines above 
$N_\nHI=10^{13.8} \cmm$, in reasonable agreement with the observations.  
}

\bigskip\bigskip
\ni If absorbers at a given surface density are conserved, with fixed comoving
space number density $n=n_0(1+z)^3$ and geometric cross-section $\Sigma$, then 
the intersection probability per unit redshift interval is
\beq
{dP\over dz}=\Sigma n {d{\ell}\over dz}=\Sigma n_0(1+z)^3 {d\ell\over dz}.
\eeq
If the Universe is cosmologically flat, the expansion rate at early epochs
is close to the Einstein-de Sitter limit, and the redshift distribution
for conserved clouds is predicted to be
\beq
{dP\over dz}\propto (1+z)^3 {d\ell\over dz}\propto (1+z)^{1/2}.
\eeq
The rate of increase of $f(N_\nHI,z)$ with $z$ in both the \lya forest and LLS 
is considerably faster than this, indicating rapid 
evolution. The mean proper distance between absorbers along the line-of-sight 
with columns greater than $N_\nHI$ is
\beq
L={d\ell\over dz} {dz\over dN}\approx {c N_\nHI^{1/2} \over 
H_0 \Omega_M^{1/2} 2 A (1+z)^{4.5}}. 
\eeq
For clouds with $N_\nHI>10^{14}\,\cmm$, this amounts to $L\sim 0.7\,h^{-1}
\Omega_M^{-1/2}\,$Mpc at $z=3$. At the same epoch, the mean proper distance
between LLS is $L\sim 30\,h^{-1}\Omega_M^{-1/2}\,$Mpc. 

\bigskip\bigskip
{\sf 
\noindent INTERGALACTIC CONTINUUM OPACITY 

Even if the bulk of the baryons in the Universe are fairly well ionized 
at all redshifts $z\lta 5$, the residual neutral hydrogen still present 
in the \lya forest clouds and Lyman-limit systems
significantly attenuates the ionizing flux from cosmological distant sources.
To quantify the degree of attenuation we have to introduce the concept of an 
effective continuum optical depth $\tau_{\rm eff}$ along the line-of-sight
to redshift $z$,
\beq
\langle e^{-\tau}\rangle\equiv e^{-\tau_{\rm eff}},
\eeq
where the average is taken over all lines-of-sight. Negleting absorption due
to helium, if we characterize the 
\lya forest clouds and LLS as a random
distribution of absorbers in column density and redshift space, then the 
effective continuum optical depth of a clumpy IGM at the observed frequency
$\nu_o$ for an observer at redshift $z_o$ is 
\begin{equation}
\tau_{\rm eff}(\nu_o,z_o,z)=\int_{z_o}^z\, dz'\int_0^{\infty}\, dN_\nHI\,
f(N_\nHI,z) (1-e^{-\tau}) \label{eq:tau}.
\end{equation}
where $\tau=N_\nHI\sigma_H(\nu)$ is the hydrogen Lyman continuum optical 
depth through an individual cloud at frequency $\nu=\nu_o(1+z)/(1+z_o)$. 
This formula can be easily understood if we consider a situation in which 
all absorbers
have the same optical depth $\tau_0$ independent of redshift, and the mean
number of systems along the path is $\Delta N=\int dzdN/dz$. In this case 
the Poissonian probability of encountering a total optical depth $k\tau_0$ 
along the line-of-sight (with $k$ integer) is $p(k\tau_0)=e^{-\Delta N} 
\Delta N^k/(\tau_0k!)$, and $\langle e^{-\tau}\rangle=e^{-k\tau_0}p(k\tau_0)=
\exp[-\Delta N(1-e^{-\tau_0})]$. 

If we extrapolate the $N_\nHI^{-1.5}$ power-law in equation (\ref{eq:dis}) to
very small and large columns, the effective optical depth becomes an analytical
function of redshift and wavelength,
\begin{equation}
\tau_{\rm eff}(\nu_o,z_o,z)={4\over 3}\sqrt{\pi\sigma_L}\, A \left({\nu_o
\over \nu_L}\right)^{-1.5} (1+z_o)^{1.5}\left[(1+z)^{1.5}-(1+z_o)^{1.5}\right].
\label{eq:t}
\end{equation}
Due to the rapid increase with lookback time of the number of
absorbers, the mean free path of photons at $912\,$\AA\ becomes so small
beyond a redshift of 2 that the radiation field is largely `local'. 
Expanding equation (\ref{eq:t}) around $z$, one gets $\tau_{\rm
eff} (\nu_L)\approx 0.36 (1+z)^2 \Delta z$. This means that at $z=3$, for
example, the mean free path for a photon near threshold is only 
$\Delta z=0.18$, and sources of ionizing radiation at higher redshifts 
are severely attenuated.
}

\bigskip\bigskip {\sf 
\ni BIBLIOGRAPHY

\ni For a recent Deuterium abundance measurement and its implications for the
baryon density parameter see

{\footnotesize Burles, S., \& Tytler, D. 1999, Ap. J., {\bf 499}, 699.}

\ni A short and hardly comprehensive list of references on issues related
to the reionization of the IGM includes  

{\footnotesize Arons, J., \& Wingert, D. W. 1972, Ap. J., {\bf 177}, 1.}

{\footnotesize Shapiro, P. R., \& Giroux, M. L. 1987, Ap. J., {\bf 321}, L107.}

{\footnotesize Meiksin, A., \& Madau, P. 1993, Ap. J., {\bf 412}, 34.}

{\footnotesize Tegmark, M., Silk, J., Rees, M. J., Blanchard, A., Abel, T.,
\& Palla, F. 1997, Ap. J., {\bf 474}, 1.}

{\footnotesize Gnedin, N. Y., \& Ostriker, J. P. 1997, Ap. J., {\bf 486}, 581.}

{\footnotesize Haiman, Z., \& Loeb, A. 1997, Ap. J., {\bf 483}, 21.}

{\footnotesize Miralda-Escude', J. 1998, Ap. J., {\bf 501}, 15.}

{\footnotesize Madau, P., Haardt, F., \& Rees, M. J. 1999, Ap. J., {\bf 514},
648.}

\ni The use of \lya resonant absorption as a sensitive probe of intergalactic
neutral hydrogen was predicted indipendently by 

{\footnotesize Gunn, J. E., \& Peterson, B. A. 1965, Ap. J., {\bf 142}, 1633.}

{\footnotesize Shklovsky, I. S. 1965, Soviet Astron., {\bf 8}, 638.}

{\footnotesize Scheuer, P. A. G. 1965, Nature, {\bf 207}, 963.}

\ni The quoted upper limit to the Gunn-Peterson optical depth at $z\approx 5$
is from 

{\footnotesize Songaila, A., Hu, E. M., Cowie, L. L., \& McMahon, R. G.
1999, Ap. J., {\bf 525}, L5.} 

\ni The high resolution quasar spectrum shown in the Figure was taken from

{\footnotesize Songaila, A. 1998, A. J., {\bf 115}, 2184,} 

\ni who also discusses the evolution of metal abundances in the \lya forest.

\ni An extensive discussion of the physics of the intergalactic medium
can be found in  

{\footnotesize Peebles, P. J. E. 1993, {\it Principles of Physical Cosmology}
(Princeton University Press), chapter 23.} 

\ni The physics of an ionized hydrogen gas is covered in

{\footnotesize Osterbrock, D. E. 1989, {\it Astrophysics of Gaseous Nebulae
and Active Galactic Nuclei} (University \par Science Books), chapters 2 and 3.}

\noindent Our present understanding of the \lya forest is summarized in 

{\footnotesize  Rauch, M. 1998, Annual Rev. of Astron. and Astrophys. 
{\bf 36}, 267.} 

\ni The use of hydrodynamic cosmological simulations to make quantitative 
prediction of the physical state of the IGM was pioneered by    

{\footnotesize Cen, R., Miralda-Escude', J., Ostriker, J. P., \& Rauch,
M. 1994, Ap. J., {\bf 437}, L9.}

\end{document}